\def \AAP #1 #2 {{\em Astron. Astrophys.\/} {\bf #1}, #2}  
\def \AAL #1 #2 {{\em Astron. Astrophys. Lett.\/} {\bf #1}, L#2}  
\def \AAR #1 #2 {{\em Astron. Astrophys. Rev.\/} {\bf #1}, #2}  
\def \AAS #1 #2 {{\em Astron. Astrophys. Suppl. Ser.\/} {\bf #1}, #2}  
\def \AJ #1 #2 {{\em Astron. J.\/} {\bf #1}, #2}  
\def \ANNREV #1 #2 {{\em Ann. Rev. Astron. Astrophys.\/} {\bf #1}, #2}  
\def \APJ #1 #2 {{\em Astrophys. J.\/} {\bf #1}, #2}  
\def \APJL #1 #2 {{\em Astrophys. J. Lett.\/} {\bf #1}, L#2}  
\def \APJS #1 #2 {{\em Astrophys. J. Suppl.\/} {\bf #1}, #2}  
\def \APSS #1 #2 {{\em Astrophys. Space Sci.\/} {\bf #1}, #2}  
\def \ASR #1 #2 {{\em Adv. Space Res.\/} {\bf #1}, #2}  
\def \MN #1 #2 {{\em Mon. Not. R. Astr. Soc.\/} {\bf #1}, #2}  
\def \PRL #1 #2 {{\em Phys. Rev. Lett.\/} {\bf #1}, #2}  
\def \NAT #1 #2 {{\em Nature\/} {\bf #1}, #2}  

\documentstyle{memsait}  
\input epsf.sty  
  
\begin{opening}  
  
\title{\LARGE\rm Cosmic Ray Acceleration at Relativistic Plasma Flows}  
  
\author{Micha{\L} Ostrowski}  
  
\institute{Obserwatorium Astronomiczne, Uniwersytet Jagiello\'nski,  
ul. Orla 171, 30-244 Krak\'ow, Poland (E-mail:  mio{@}oa.uj.edu.pl)}  
  
\date{} % DO NOT INSERT ANY DATE HERE !!!  
\end{opening}  
  
\begin{document}  
  
\oddpagefooter{}{}{} % LEAVE AS IT IS !  
\evenpagefooter{}{}{} % LEAVE AS IT IS !  
\   
\bigskip  
  
\begin{abstract}  
Theoretical concepts of cosmic ray particle acceleration at 
relativistic plasma flows -- shocks and shear layers -- are reviewed. We 
begin with a discussion of mildly relativistic shock waves. The role of 
oblique field configurations and field perturbations in forming the 
particle energy spectrum and changing the acceleration time scale is 
considered. Then, we report on two interesting attempts to consider 
particle acceleration at ultra-relativistic shocks. Finally, in contrast 
to the compressive shock discontinuities, we discuss the acceleration 
processes acting in the boundary layer at the tangential velocity 
transition. The second-order Fermi acceleration as well as the cosmic 
ray `viscous' acceleration provide the  mechanisms generating energetic 
particles there. \end{abstract}

\section{Introduction}  
  
Relativistic plasma flows are detected or postulated to exist in a  
number of astrophysical objects, ranging from a mildly relativistic jet of  
SS433, through the-Lorentz-factor-of-a-few jets in AGNs and galactic  
`mini-quasars', up to ultra-relativistic outflows in sources of gamma  
ray bursts and, possibly, in pulsar winds. As nearly all such objects  
are efficient emitters of synchrotron radiation and/or high energy  
photons requiring the existence of energetic particles, our attempts to  
understand the processes generating cosmic ray particles are essential for  
understanding the fascinating phenomena observed. Below we will discuss  
the work carried out  in order to understand the cosmic ray acceleration processes acting  
at relativistic flow discontinuities -- shocks and shear layers. The  
present review is an updated version of my earlier presentations  
(Ostrowski 1996, 1997), also including the discussion of the acceleration  
at the ultra-relativistic shock by Bednarz \&  Ostrowski (1998) and the  
detailed physical model for the pulsar wind terminal shock by Arons and  
collaborators (cf. Arons 1996).  
  
\section{Particle acceleration at non-relativistic shock waves}  
  
Processes of the first-order particle acceleration at non-relativistic shock  
waves were widely discussed by a number of authors during the last two  
decades (for review, see, e.g. Drury 1983, Blandford \& Eichler 1987,  
Berezhko et al. 1988, Jones \& Ellison 1991). Below, we review  the basic  
physical picture and some important results obtained within this theory  
{\it for test particles}, to be later compared with the results obtained  
for relativistic shocks.  
  
The simple description of the acceleration process preferred by us 
consists of  considering two plasma rest frames, the {\it upstream frame}  
and the {\it downstream one}. We use indices `$1$' or `$2$' to indicate  
quantities measured in the upstream or the downstream frame  
respectively. If one neglects the second-order Fermi acceleration, the  
particle energy is a constant of motion in any of these plasma rest  
frames and energy changes occur when the particle momentum is  
Lorentz-transformed at each crossing of the shock. In the case of {\it  
parallel} shock, with the mean magnetic field parallel to the shock  
normal, the acceleration of an individual particle is due to the   
consecutive shock crossings by the diffusive wandering particle. Each  
{\it upstream-downstream-upstream} diffusive loop results in a small  
increment of particle momentum, $\Delta p \propto p \cdot (U_1-U_2)/v$,  
where $v$ is the particle velocity and $U_i$ is the shock velocity in  
the respective $i = 1$ or $2$ frame, $U_1 \ll v $. In oblique  
shocks, the particle helical trajectories can cross the shock surface a  
number of times at any individual shock transition or reflection.  
  
The most interesting feature of the first-order Fermi acceleration at a  
non-relativistic plane-parallel shock wave is the independence of the {\it  
test-particle stationary} particle energy spectrum from the background  
conditions near the shock, including the mean magnetic field  
configuration and the spectrum of MHD turbulence. The main reason behind  
that is a nearly-isotropic form of the particle momentum distribution at  
the shock. If a sufficient amount of scattering occurs near the shock,  
this condition always holds for the shock velocity along the upstream  
magnetic field $U_{B,1} \equiv U_1 / \cos \Psi_1 \ll v$ ($\Psi_1$ - the  
upstream magnetic field inclination to the shock normal). Independently  
of the field inclination at the shock, the particle density is  
continuous across it and the spectral index for the phase-space  
distribution function, $\alpha$, is given exclusively in the terms of  
a single parameter -- the shock compression ratio $R$:  
  
$$\alpha = {3R \over R-1}  \qquad   . \eqno(2.1) $$  
  
\noindent  
Because of the isotropic form of the particle distribution function, the  
spatial diffusion equation has become a widely used mathematical tool  
for describing particle transport and acceleration processes in  
non-relativistic flows. Thus the characteristic acceleration time scale  
at the parallel ($\Psi_1=0$) shock is  
  
$$T_{acc} = {3 \over U_1-U_2}\, \left\{ {\kappa_1 \over U_1} + {\kappa_2  
\over U_2} \right\} \qquad , \eqno(2.2) $$  
  
\noindent  
where $\kappa_i \equiv \kappa_{\parallel ,i}$ is the respective particle  
spatial diffusion coefficient along the magnetic field, as discussed by  
e.g. Lagage \& Cesarsky (1983). Ostrowski (1988a; see also Bednarz \&  
Ostrowski 1996) derived an analogous expression for shocks with oblique  
magnetic fields and small amplitude magnetic field perturbations. For a  
negligible cross-field diffusion and for $U_{B,1} \ll c$ it can be  
written in essentially the same form as the one given in Eq.~(2.2), with  
all quantities taken as the normal ($n$) ones with respect to the shock  
($\kappa_{n,i}$ for $\kappa_i$ ($i$ = $1$, $2$)). As $\kappa_n <  
\kappa_\parallel$, the oblique shocks may be more rapid accelerators when   
compared to the parallel shocks.  
  
Not discussed here non-linear and time dependent effects, inclusion of  
additional energy losses and gains, etc., make the physics of the  
acceleration more intricate, allowing e.g. for non-power-low and/or  
non-stationary particle distributions.  
  
\section{Cosmic ray acceleration at relativistic shock waves}  
  
\subsection{The Fokker-Planck description of the acceleration process}  
  
In the case of the shock velocity (or its projection $U_{B,1}$) reaching  
values comparable to the light velocity, the particle distribution at  
the shock becomes anisotropic. This fact complicates to a great extent  
both the physical picture and the mathematical description of particle  
acceleration. The first attempt to consider the acceleration process at  
the relativistic shock was presented in 1981 by Peacock (see also Webb  
1985); however, no consistent theory was proposed until a paper of Kirk  
\& Schneider (1987a; see also Kirk 1988) appeared. Those authors  
considered the stationary solutions of the relativistic Fokker-Planck  
equation for particle pitch-angle diffusion for the case of the parallel  
shock wave. In the situation with the gyro-phase averaged distribution  
$f(p, \mu, z)$, which depends only on the unique spatial co-ordinate $z$  
along the shock velocity, and with $\mu$ being the pitch-angle cosine,  
the equation takes the form:  
  
$$\Gamma ( U + v \mu ) {\partial f \over \partial z} = C(f) + S \qquad ,  
 \eqno(3.1)  $$  
  
\noindent  
where $\Gamma \equiv 1/\sqrt{1-U^2}$ is the flow Lorentz factor, $C(f)$  
is the collision operator and $S$ is the source function. In the  
presented approach, the spatial co-ordinates are measured in the shock  
rest frame, while the particle momentum co-ordinates and the collision  
operator are given in the respective plasma rest frame. For the applied  
pitch-angle diffusion operator, $C = \partial / \partial \mu (D_{\mu  
\mu} \partial f / \partial \mu)$, they generalised the diffusive  
approach to higher order terms in particle distribution anisotropy and  
constructed general solutions at both sides of the shock which involved  
solutions of the eigenvalue problem. By matching two solutions at the  
shock, the spectral index of the resulting power-law particle  
distribution can be found by taking into account a sufficiently large  
number of eigenfunctions. The same procedure yields the particle angular  
distribution and the spatial density distribution. The low-order  
truncation in this approach corresponds to the standard diffusion  
approximation and to a somewhat more general method described by  
Peacock. The above analytic approach (or the `semi-analytic' one, as the  
mentioned matching of two series involves numerical fitting of the  
respective coefficients) was verified by Kirk \& Schneider (1987b) by  
the method of particle Monte Carlo simulations.

\begin{figure}[hbt]  
\vspace{70mm}
\includegraphics{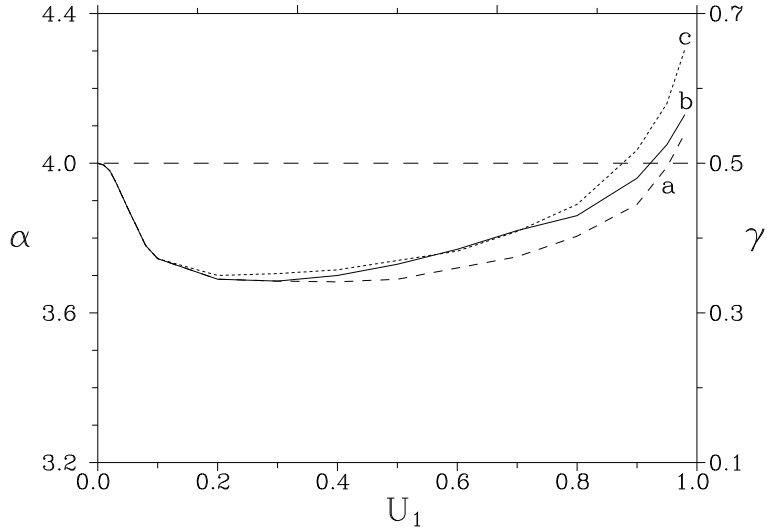}   
\vspace{0mm}
\caption{ 
The particle spectral indices $\alpha$ at parallel shock  
waves propagating in the cold ($e$, $p$) plasma versus the shock  
velocity $U_1$ (Heavens \& Drury 1988). On the right vertical axis the  
respective synchrotron spectral index $\gamma$ is given. Using the solid  
line (b) and the dashed line (a) we show indices for two choices of the  
turbulence spectrum. The dashed line (c) gives the spectral index  
derived from Eq.~2.1. The horizontal line $\alpha = 4.0$ is given for  
the reference.} \end{figure}  
  
An application of this approach to more realistic conditions -- but  
still for parallel shocks -- was presented by Heavens \& Drury (1988),  
who investigated the fluid dynamics of relativistic shocks (cf. also  
Ellison \& Reynolds 1991) and used the results to calculate spectral  
indices for accelerated particles (Fig.~1). They considered the shock  
wave propagating into electron-proton or electron-positron plasma, and  
performed calculations using the analytic method of Kirk \& Schneider  
for two different power spectra for the scattering MHD waves. In  
contrast to the non-relativistic case, they found (see also Kirk 1988)  
that the particle spectral index depends on the form of the wave  
spectrum.  The unexpected fact was noted that the non-relativistic  
expression (2.1) provided a quite reasonable approximation to the actual  
spectral index.  
  
\begin{figure}[hbt]  
\vspace{70mm}
\includegraphics{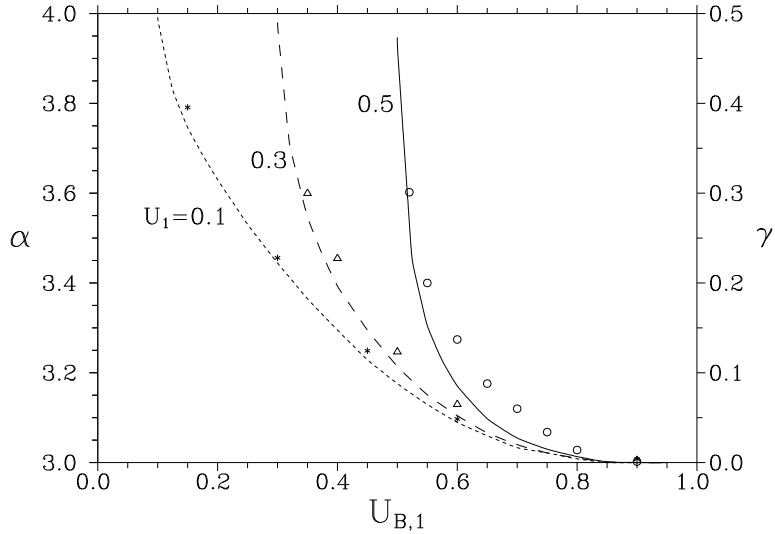}   
\vspace{0mm}
\caption{ 
Spectral indices $\alpha$ of particles accelerated at oblique shocks  
versus shock velocity projected at the mean magnetic field, $U_{B,1}$.  
On the right the respective synchrotron spectral index $\gamma$ is  
given. The shock velocities $U_1$ are given near the respective curves  
taken from Kirk \& Heavens (1989). The points were taken from  
simulations deriving explicitly the details of particle-shock  
interactions (Ostrowski 1991a). The results are presented for  
compression $R = 4$.} \end{figure}  
  
A substantial progress in understanding the acceleration process in the  
presence of highly anisotropic particle distributions is due to the work  
of Kirk \& Heavens (1989; see also Ostrowski 1991a and Ballard \&  
Heavens 1991), who considered particle acceleration at {\it subluminal}  
($U_{B,1} < c$) relativistic shocks with oblique magnetic fields. They  
assumed the magnetic momentum conservation, $p_\perp^2/B = const$, at  
particle interaction with the shock and applied the Fokker-Planck  
equation discussed above to describe particle transport along the field  
lines outside the shock, while excluding  the possibility of cross-field  
diffusion. In the cases when $U_{B,1}$ reached relativistic values, they  
derived very flat energy spectra with $\gamma \approx 0$ at $U_{B,1}  
\approx 1$ (Fig.~2). In such conditions, the particle density in front of the  
shock can substantially -- even by a few orders of magnitude -- exceed  
the downstream density (see the curve denoted `-8.9' at Fig.~3).  
Creating flat spectra and great density contrasts is due to the effective  
reflections of anisotropically distributed upstream particles from the  
region of compressed magnetic field downstream of the shock. However, the  
conditions leading to very flat spectra are supposed to be accompanied  
by processes -- like a large amplitude wave generation upstream of the  
shock -- leading to spectrum steepening (cf. Sec.~3.2).  
  
As stressed by Begelman \& Kirk (1990), in relativistic shocks one can  
often find the {\it superluminal} conditions with $U_{B,1} > c$, where  
the above presented approach is no longer valid. Then, it is not  
possible to reflect upstream particles from the shock and to transmit  
downstream particles into the upstream region. In effect, only a single  
transmission of upstream particles re-shapes the original distribution  
by shifting particle energies to larger values. The energy gains in such  
a process, involving a highly anisotropic particle distribution, can be  
quite significant, exceeding the value expected for the adiabatic  
compression.  
  
The approach proposed by Kirk \& Schneider (1987a) and Kirk \& Heavens  
(1989), and the derivations of Begelman \& Kirk (1990) are valid only in  
case of weakly perturbed magnetic fields. However, in the  
efficiently accelerating shocks one may expect  large amplitude waves  
to be present, when both the Fokker-Planck approach is no longer valid  
and the magnetic momentum conservation no longer holds for oblique  
shocks. In such a case, numerical methods have to be used.  
  
\subsection{Particle acceleration in the presence of large amplitude  
magnetic field perturbations}  
  
\begin{figure}[hbt]  
\vspace{76mm}
\includegraphics{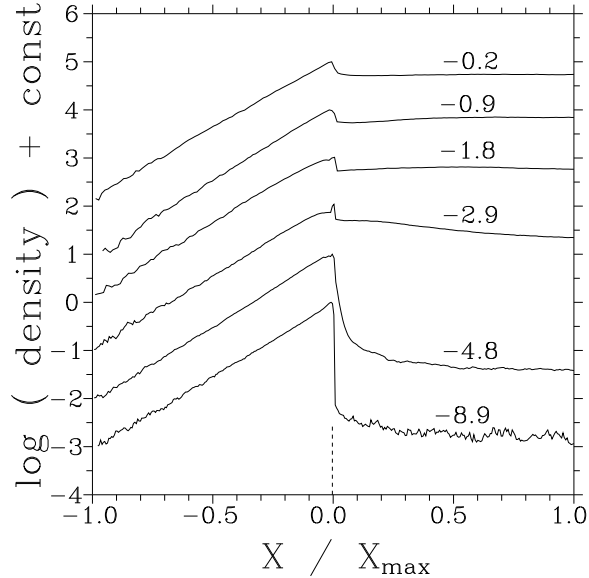}   
\vspace{0mm}
\caption{ 
The energetic particle density across the relativistic shock  
with an oblique magnetic field (Ostrowski 1991b). The shock with $U_1 =  
0.5$, $R = 5.11$ and $\psi_1 = 55^o$ is considered. The curves for  
different perturbation amplitudes are characterized with the value $\log  
\kappa_\perp / \kappa_\parallel$ given near the curve. The data are  
vertically shifted for picture clarity. The value $X_{max}$ is the  
distance from the shock at which the upstream particle density decreases  
to $10^{-3}$ part of the shock value.} \end{figure}  
  
The first attempt to consider the acceleration process at parallel shock  
wave propagating in a turbulent medium was presented by Kirk \&  
Schneider (1988), who included into Eq.~3.1 the Boltzmann collision  
operator describing the large angle scattering. By solving the resulting  
integro-differential equation they demonstrated the hardening of the  
particle spectrum due to increasing contribution of the large-angle  
scattering. The reason for such a spectral change is the additional  
isotropization of particles interacting with the shock, leading to an   
increase in the particle mean energy gain. In oblique shocks, this simplified  
approach cannot be used because the character of individual  
particle-shock interaction -- reflection and transmission  
characteristics -- depends on the magnetic field perturbations. Let us  
additionally note that application of the point-like large-angle  
scattering model in relativistic shocks does not provide a viable  
physical representation of the scattering at MHD waves (Bednarz \&  
Ostrowski 1996).  
  
\begin{figure}[hbt]  
\vspace{71mm}
\includegraphics{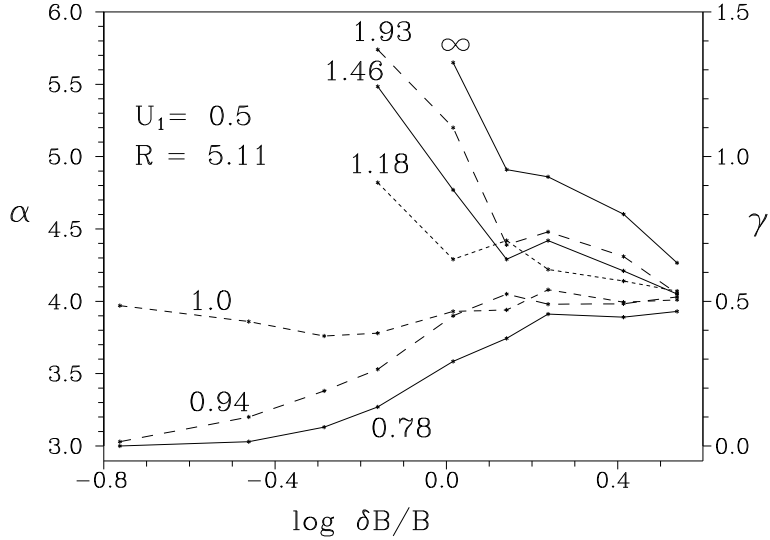}   
\vspace{0mm}
\caption{ 
Spectral indices for oblique relativistic shocks versus  
perturbation amplitude $\delta B/B$ (Ostrowski 1993). Different field  
inclinations are characterized by the values of $U_{B,1}$ given near the  
respective results, $U_{B,1} < 1$ for subluminal shocks and $U_{B,1} \ge  
1$ for superluminal ones. Absence of data for small field amplitudes in  
superluminal shocks is due to extremely steep power law spectra  
occurring in these conditions (cf. Begelman \&  Kirk 1990). Decreasing  
the field inclination $\Psi_1 \to 0$ (i.e to the parallel shock with  
$U_{B,1} = U_1$) gives spectral indices more and more similar to a  
constant line $\alpha = 3.72$,  not shown here  
for picture clarity (cf. Fig-s~1,2).} \end{figure}  
  
To handle the problem of the particle spectrum in a wide range of background  
conditions, the Monte Carlo particle simulations were proposed (Kirk \&  
Schneider 1987b; Ellison et al. 1990; Ostrowski 1991a, 1993; Ballard \&  
Heavens 1992, Naito \& Takahara 1995, Bednarz \&  Ostrowski 1996, 1998).  
At first, let us consider subluminal shocks. The field perturbations  
influence the acceleration process in various ways. As they enable the  
particle cross field diffusion, a modification (decrease) of the  
downstream particle's escape probability may occur. This factor tends to  
harden the spectrum. Next, the perturbations decrease particle  
anisotropy, leading to an increase of the mean energy gain of reflected  
upstream particles, but -- what is more important for oblique shocks --  
this also increases the particle upstream-downstream transmission  
probability due to less efficient reflections, enabling them to escape  
from further acceleration. The third factor is due to perturbing  
particle trajectory during an individual interaction with the shock  
discontinuity and breakdown of the approximate conservation of  
$p_\perp^2/B$. Because reflecting a particle from the shock requires a  
fine tuning of the particle trajectory with respect to the shock surface,  
even small amplitude perturbations can decrease the reflection  
probability in a substantial way. Simulations show (see Fig.~4 for  
$U_{B,1} < 1.0$) that -- until the wave amplitude becomes very large --  
the factors leading to efficient particle escape dominate with the  
resulting steepening of the spectrum to $\gamma \sim 0.5$ -- $0.8$, and  
the increased downstream transmission probability lowers the cosmic ray  
density contrast across the shock (Fig.~3).  
  
In parallel shock waves propagating in a highly turbulent medium, the  
effects discovered for oblique shocks can also manifest their  presence  
because of the {\it local} perturbed magnetic field compression at the  
shock. The problem was considered using the technique of particle  
simulations by Ballard \& Heavens (1992; cf. Ostrowski 1988b for  
non-relativistic shock). They showed a possibility of having a very steep  
spectrum in this case, with the spectral index growing from $\gamma \sim  
0.6$ at medium relativistic velocities up to nearly $2.0$ at $U_1 =  
0.98$. These results apparently do not correspond to the 
large-perturbation-amplitude limit of Ostrowski's (1993; see the  
discussion therein) simulations for oblique shocks and the analytic  
results of Heavens \& Drury (1988).  
  
For large amplitude magnetic field perturbations the acceleration  
process in superluminal shocks can lead to the power-law particle  
spectrum formation, against the statements of Begelman \& Kirk (1990)  
valid at small wave amplitudes only. Such  a general case was discussed by  
Ostrowski (1993; see Fig.~4 for $U_{B,1} \ge 1$) and by Bednarz \&  
Ostrowski (1996, 1998).  
  
\subsection{The acceleration time scale}  
  
\begin{figure}[hbt]  
\vspace{72mm}
\includegraphics{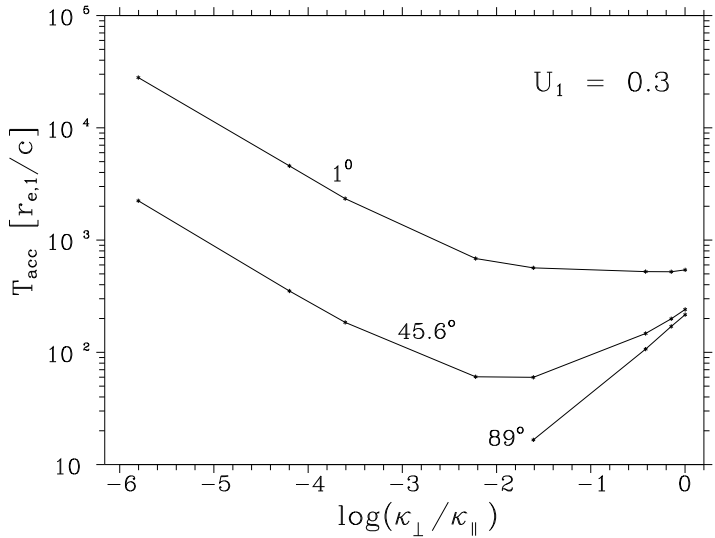}   
\vspace{0mm}
\caption{ 
The acceleration time $T_{acc}$ versus the level of particle  
scattering measured by the ratio of $\kappa_\perp / \kappa_\parallel$(  
Bednarz \& Ostrowski 1996). We present results for three  
values of the magnetic field inclination: a.) parallel shock ($\psi_1 =  
1^\circ$), b.) a subluminal shock with $\psi_1 = 45.6^\circ $ and c.) a  
superluminal shock with $\psi_1 = 89^\circ $. $r_{e,1}$ is the particle   
gyroradius in the effective (including  
perturbations) upstream magnetic field .}  
\end{figure}  
  
\noindent  
The shock waves propagating with relativistic velocities also raise  
interesting questions pertaining to the cosmic ray acceleration time  
scale, $T_{acc}$. A simple comparison to  non-relativistic values  
shows that $T_{acc}$ relatively decreases with increasing shock velocity  
for parallel (Quenby \& Lieu 1989; Ellison et al. 1990) and oblique  
(Takahara \& Terasawa 1990; Newman et al. 1992; Lieu et al. 1994; Quenby  
\& Drolias 1995; Naito \& Takahara 1995) shocks. However, the numerical  
approaches used there, based on assuming particle isotropization for all  
scatterings, neglect or underestimate a significant factor affecting the  
acceleration process -- the particle anisotropy. Ellison et al. (1990)  
and Naito \& Takahara (1995) also included the more realistic, in our  
opinion, derivations involving the pitch-angle diffusion approach. The  
calculations of Ellison et al. for parallel shocks show similar results  
to those they obtained for large amplitude scattering. For the shock with  
velocity $0.98\,c$ the acceleration time scale is reduced by the factor  
$\sim 3$ with respect to the non-relativistic formula of Eq.~2.2~. Naito  
\& Takahara considered shocks with oblique magnetic fields. They  
confirmed the reduction of the acceleration time scale with an increasing  
inclination of the magnetic field, derived earlier for non-relativistic  
shocks. However, their approach neglected effects of particle cross  
field diffusion and assumed the adiabatic invariant conservation in  
particle interactions with the shock, thus limiting the validity of their  
results to a small amplitude turbulence near the shock.  
  
\begin{figure}[hbt]  
\vspace{72mm}
\includegraphics{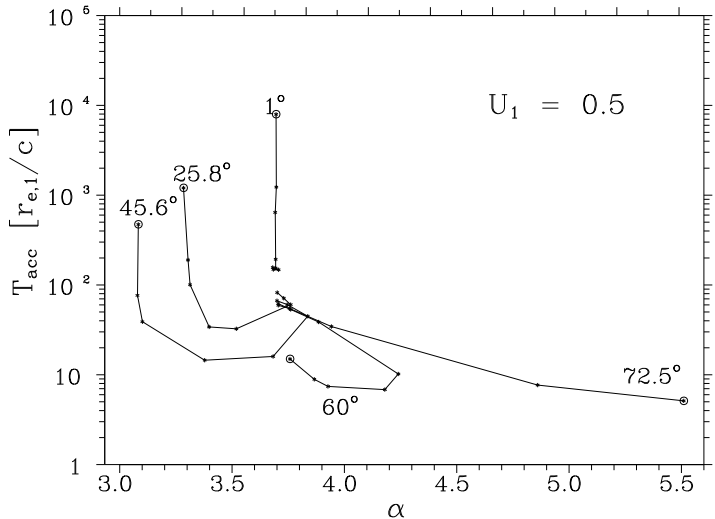}   
\vspace{0mm}
\caption{ 
The relation of $T_{acc}$ versus the particle spectral index  
$\alpha$ at different magnetic field inclinations $\psi_1$ given near  
the respective curves. The {\it minimum} value of the model parameter  
$\kappa_\perp/\kappa_\|$ occurs at the encircled point of each curve and  
the wave amplitude monotonously increases along each curve up to $\delta  
B \sim B$; $r_{e,1}$ -- see Fig.~5.} \end{figure}  
  
A wider discussion of the acceleration time scale is presented by  
Bednarz \& Ostrowski (1996), who apply numerical simulations involving  
the small angle particle momentum scattering. The approach is also believed  
to provide a reasonable description of particle transport in the  
presence of large $\delta B$, and thus to enable modelling of the effects  
of cross-field diffusion. The resulting values (Fig-s~5, 6) are given in  
the shock {\it normal} rest frame (cf. Begelman \&  Kirk 1990). In  
parallel ($\Psi_1 = 1^\circ$) shocks $T_{acc}$ diminishes with the  
growing perturbation amplitude and  shock velocity $U_1$. However, it  
is approximately constant for a given value of $U_1$ if we use the  
formal diffusive time scale, $\kappa_1/(U_1c) + \kappa_2/(U_2c)$, as the  
time unit. A new feature discovered in oblique shocks is that due to the  
cross-field diffusion $T_{acc}$ can change with $\delta B$ in a  
non-monotonic way (Fig.~5). The acceleration process leading to the  
power-law spectrum is possible in superluminal shocks only in the  
presence of large amplitude turbulence. Then, in contrast to the  
quasi-parallel shocks, $T_{acc}$ increases with increasing $\delta B$.  
In the considered cases with the oblique field configurations one may  
note a possibility to have an extremely short acceleration time scale  
comparable to the particle gyroperiod in the magnetic field upstream of the  
shock. A coupling between the acceleration time scale and the particle  
spectral index is presented in Fig.~6. One should note that the form of  
involved relation is contingent to a great extent on the magnetic field  
configuration.  
  
\subsection{Energy spectra of cosmic rays accelerated at large  
Lorentz-factor shocks}  
  
\noindent  
The main difficulty in modelling the acceleration process at shocks with  
large Lorentz factors $\Gamma$ is the fact that the involved particle  
distributions are extremely anisotro\-pic in the shock, with the  
particle angular distribution opening angles $\sim \Gamma^{-1}$ in the  
upstream plasma rest frame. When transmitted downstream of the shock,  
particles have a limited chance to be scattered so efficiently as to reach  
the shock again, but the energy gain of any such `successful' particle  
can be comparable to, but not much larger than it's original  
energy\footnote{It is impossible to reflect a charged particle from the  
large-$\Gamma$ shock if a magnetic field is present upstream of it. The  
shock will always overtake the upstream escaping particle. The  
reflections with relative energy gains $\sim \Gamma^2$ are in principle  
possible from the sides of jets (see below).}. In the simulations of  
Bednarz \& Ostrowski (1998) a hybrid  Monte Carlo method involving small  
amplitude pitch-angle scattering is applied for particle transport near  
the shock with $\Gamma$ in the range $3$ -- $243$. The same scattering  
conditions upstream and downstream of the shock (the same $\kappa_\perp$  
and $\kappa_\|$ in the units of $r_g c$, where $r_g$ is the particle  
gyration radius in the unperturbed background magnetic field) were  
assumed. A few configurations of the upstream magnetic field, with  
inclinations respective to the shock normal being $\psi = 0^\circ$,  
$10^\circ$, $20^\circ$, $30^\circ$, $60^\circ$ and $90^\circ$ were  
considered. The first case represents the parallel shock, the second is  
for the oblique -- subluminal at $\Gamma = 3$ and superluminal at larger  
$\Gamma$ -- shock, and the larger $\psi$ are for superluminal  
perpendicular shocks at all velocities. The downstream magnetic field is  
derived for the relativistic shock with the compression $R$ obtained  
with the formulae of Heavens and Drury (1988) for a cold ($e$, $p$)  
plasma (however, at so high $\Gamma$ it weakly depends on this  
particular choice).  
  
\begin{figure}[hbt]  
\vspace{70mm}
\includegraphics{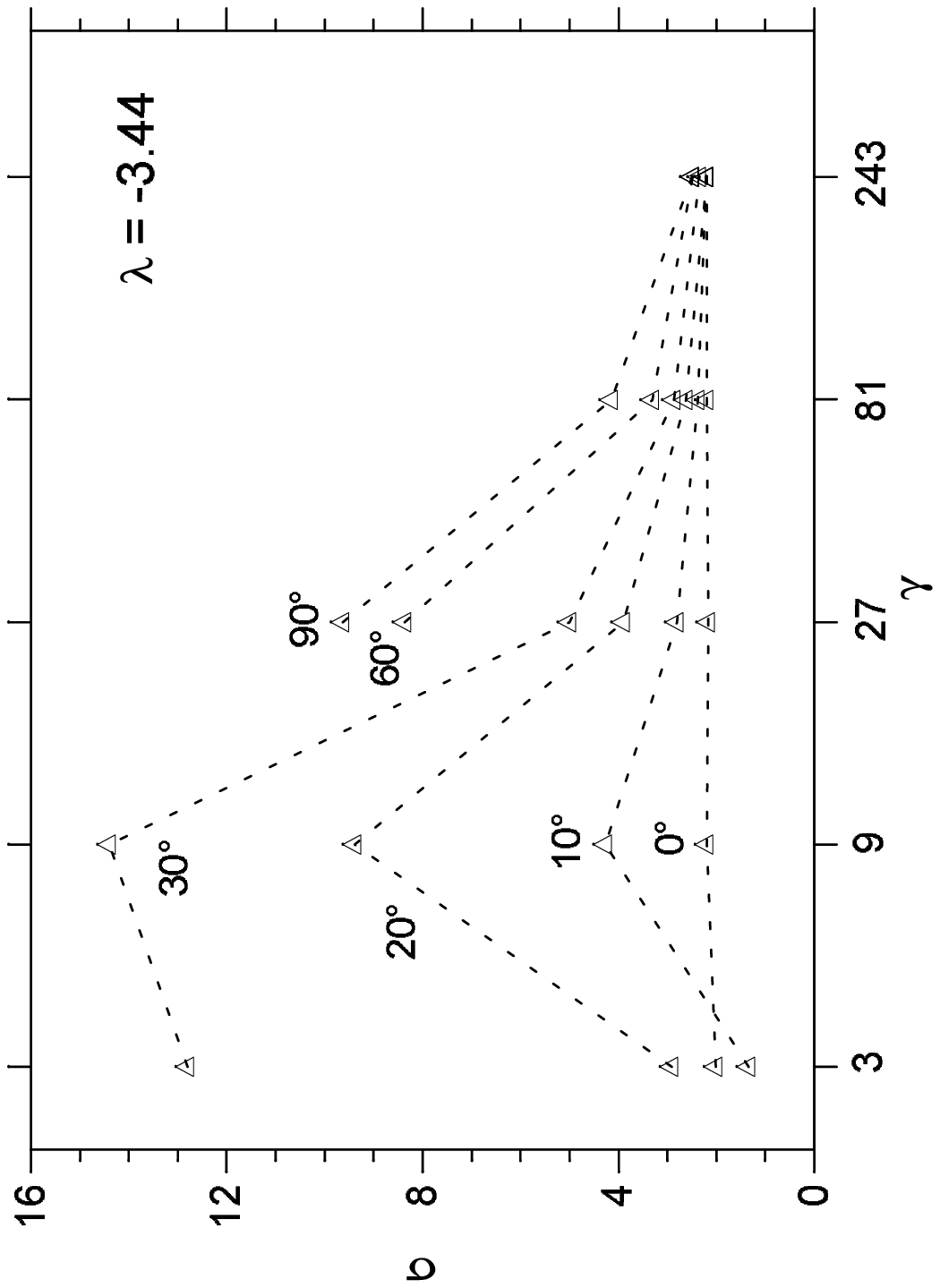}   
\vspace{0mm}
\caption{ 
The simulated spectral indices $\sigma$ ($\sigma \equiv  
\alpha - 2$) versus $\Gamma$. Results for a given $\psi$ are joined with  
dashed lines; the respective value of $\psi$ is given near each curve.  
Increasing of parameter $\lambda$ results in shifting curves toward the  
stable $\Psi = 0^\circ$ results.} \end{figure}  
  
Results for varying $\psi$ at some intermediate $\lambda \equiv  
\kappa_\perp / \kappa_\|$ are presented at Fig.~7. For the parallel  
shock ($ \psi = 0^\circ $) the amount of scattering does not influence  
the spectral index and for growing $\Gamma$ it approaches $\sigma_\infty  
\simeq 2.2$. Essentially the same limiting value was anticipated for the  
large-$\Gamma$ parallel shocks by Heavens and Drury (1988). The results  
for $\psi = 10^\circ$ are for superluminal shocks if $\Gamma > 5.75$. In  
this case, when going from the `slow' $\Gamma = 3$ shocks to the higher  
$\Gamma$ ones, at first the spectrum inclination increases ($\sigma$  
grows), but at large $\Gamma$ the spectrum flattens to approach the  
asymptotic value close to 2.2~. The spectrum steepening phase -- usually  
interpreted as an energy cut-off -- is more pronounced for small  
$\kappa_\perp / \kappa_\|$~, but even at very low turbulence levels the  
final range of spectrum flattening is observed. For larger $\psi$  
the situation does not change, but the phase of spectrum steepening is  
wider, involves larger values of $\sigma$ and starts at smaller  
velocities. In the case of large spectral indices occurring in the  
steepening phase the main factor increasing the particle energy density  
is a non-adiabatic compression in the shock (Begelman \& Kirk 1990).  
  
The inspection of particle trajectories reveals a simple picture of  
acceleration. Cosmic ray particles are wandering in the downstream  
region with the shock wave moving away with the mildly relativistic  
velocity $ \approx c/3$. Some of these particles succeed to reach the  
shock, but then they remain in the upstream region for a very short time  
-- being very close to the shock -- due to large shock velocity $\approx  
c$. This scenario is essentially equivalent to the picture involving  
downstream particles reflecting in a non-elastic way from the receding  
wall of the shock. For large-$\Gamma$ shocks any particle crossing the  
shock upstream has a momentum vector nearly parallel to the shock  
normal, the momentum inclination must be smaller than $\theta_{max} =  
\arccos (U_1) \ll 1$. If the scattering or a movement along the curved  
trajectory increases this inclination above $\theta_{max}$ the particle  
tends to re-cross the shock downstream. One should note that even a tiny  
-- comparable to $\theta_{max}$ -- angular deviation in the upstream  
plasma ($\Delta \theta_{U}$) can lead to large  
angular deviation for $\Gamma \gg 1$ as observed in the downstream rest frame. The  
phenomenon of decreasing $\sigma$ to $\sigma_\infty$ at constant  
$\lambda$ and for growing $\Gamma$ results from slower diminishing of  
the part of $\Delta \theta_{U}$ caused by scattering  in comparison to  
the one arising due to trajectory curvature in the uniform field  
component \footnote{This way the magnetic field structure defined by  
$\psi$ becomes unimportant. One should also note that the downstream  
field inclination approaches $90^\circ$ if $\psi\neq 0$ and $\Gamma \to  
\infty$.}. As a result the particles crossing the shock downstream are  
scattered in a wide angular range with respect to the shock normal,  
providing some particles with parameters allowing for re-crossing  
upstream of the shock even for the perpendicular magnetic field  
configuration. An interesting finding, not fully explained with such  
simple arguments is the fact that the number of such particles  
re-crossing the shock becomes nearly independent of the magnetic field  
inclination and the turbulence amplitude. It is also observed that when  
approaching the limiting value of the spectral index the mean particle  
energy gain $<\Delta E/E>$ in the cycle `upstream-downstream-upstream'  
reaches a value close (slightly above) to $1.0$, much smaller than the  
factor $\sim \Gamma^2$ expected for a model involving large angle  
point-like scattering.  
  
\subsection{Acceleration at the ultra-relativistic shock near the  
Crab Pulsar}  
  
As discussed above, the details of the acceleration process can  
substantially modify particle spectra in relativistic shocks. Thus the 
knowledge of some details of the shock transition deserves  the   
effort to make the model more specific. Such an approach was presented by  
Arons and collaborators (e.g. Hoshino et al. 1992, Gallant \& Arons  
1994; for review Arons 1996), who considered acceleration at the  
ultra-relativistic shock formed in the wind outflow of the (e$^+$,e$^-$)  
pair plasma containing heavy nuclei and being permeated by the weak  
magnetic field oriented perpendicular to the flow direction, i.e. in a  
model wind for the Crab Pulsar. In the large Lorentz factor wind, the ram  
pressure of nuclei dominates over the ram pressure of the pair plasma,  
and both these pressures are much larger than the magnetic field  
pressure.  
  
At the collisionless shock, the pairs' bulk velocity is isotropiz\-ed  
much more efficiently, leaving nuclei penetrating the downstream region  
as a particle beam. This process generates an electric field in the shock  
and -- due to the ion distribution anisotropy -- generates long  
electromagnetic plasma waves. Damping of such waves by pairs accelerates  
some of electrons/positrons to energies allowing for the creation of the  
observed TeV photons and synchrotron radiation up to several tens of  
MeV. In this model, gyroradia of nuclei are comparable to the radial size  
of the shock ($\approx 0.1$ pc) and this component is suspected to be  
responsible for optical wisps observed in the Crab (Gallant \& Arons  
1994). The work mentioned here is based on the results of numerical plasma  
simulations of the ultra-relativistic collisionless shock. If  
this approach points into the right physics, then the presented work  
could be evaluated as the only  realistic model of the acceleration  
process for relativistic shocks available in  literature.  
  
\section{Energetic particle acceleration at a relativistic shear layer}  
       
There is a list of observations pointing to the fact that  
acceleration processes acting e.g. in AGNs central sources and in shocks  
formed in jets are not always able to explain the observed high  
energy electrons radiating away from the centre/shock. Among a few  
proposed possibilities explaining these data the relatively natural but  
unexplored is the one involving particle acceleration at a tangential  
velocity `jump' or a shear layer at the interface between the jet and  
the surrounding medium (`cocoon'). To date the knowledge of physical  
conditions within such layers is very limited and only rough estimates  
for the considered acceleration processes are possible. Within the {\it  
subsonic} turbulent layer with a non-vanishing small velocity shear the  
ordinary second-order Fermi acceleration, as well as the  
process of `viscous' particle acceleration (cf. the review by Berezhko  
(1990) of the work done in early 80-th; also Earl et al. 1988, Ostrowski  
1990, 1998)can take place. A mean particle energy gain in the later process scales as  
  
$$ <\Delta E> \, \propto \left( { <\Delta U> \over c } \right)^2 \qquad ,   
\eqno(4.1)$$   
  
\noindent  
where $< \Delta U >$ is the mean velocity difference between the  
`successive scattering acts'. It is proportional to the mean free path  
normal to the shear layer, $\lambda_n$, times the mean flow velocity  
gradient in this direction $ \nabla_n \cdot \vec{U} $. With $d$ denoting  
the shear layer thickness this gradient can be estimated as $| \nabla_n  
\cdot \vec{U} | \approx U/d$. Because the acceleration rate in the Fermi  
II process is $\propto (V / c)^2$ ($V \approx V_A$ is the wave velocity,  
$V_A$ -- the Alfv\'en velocity), the relative importance of both  
processes is given by a factor  
  
$$\left( {\lambda_n \over d} {U \over V} \right)^2 \qquad . \eqno(4.2)$$  
  
\noindent  
The relative efficiency of the viscous acceleration grows with  
$\lambda_n$ and in the formal limit of $\lambda_n \approx d$ -- outside  
the equation (4.2) validity range -- it dominates over the Fermi  
acceleration to a  large extent. Because accelerated particles can  
escape from the accelerating layer only due to a relatively inefficient  
radial diffusion, the resulting particle spectra should be very flat up  
to the high energy cut-off, but the exact form of the spectrum depends  
on several unknown physical parameters of the boundary layer. More solid  
information can be obtained about generated spectra of {\it `very high'}  
energy particles, where a particle belongs to this class if $\lambda_n  
\ge d$ (or $r_g > d$). A few results for such particles accelerated in  
the absence of radiation losses are presented below.  
     
\subsection{Acceleration of  very high energy particles: terminal  
shock versus the jet's side boundary}  
     
For sufficiently energetic particles both the jet terminal shock and the  
transition layer between the jet and the surrounding medium can be  
approximated as surfaces of discontinuous velocity change. The later  
tangential discontinuity can be an efficient cosmic ray acceleration  
site if the considered velocity difference is relativistic and the  
sufficient amount of turbulence on its both sides is present (Berezhko  
1990, Ostrowski 1990). On average, at a single boundary crossing a  
particle gains  
  
$$<\Delta E> \, = \, \rho_e \, (\Gamma-1) \, E_0 \qquad , \eqno(4.3)$$  
  
\noindent  
where $\Gamma$ is the jet Lorentz factor and the numerical factor  
$\rho_e$ depends on the particles' anisotropy at the discontinuity. Particle  
simulations described by Ostrowski (1990),  in the presence of efficient  
particle scattering, give $\rho_e$ as a substantial fraction of unity.  
Let us also note that in the case of a non-relativistic velocity jump, $U  
\ll c$, the acceleration process is of the second-order in $U/c$.  
  
\begin{figure}[hbt]  
\vspace{55mm}
\includegraphics{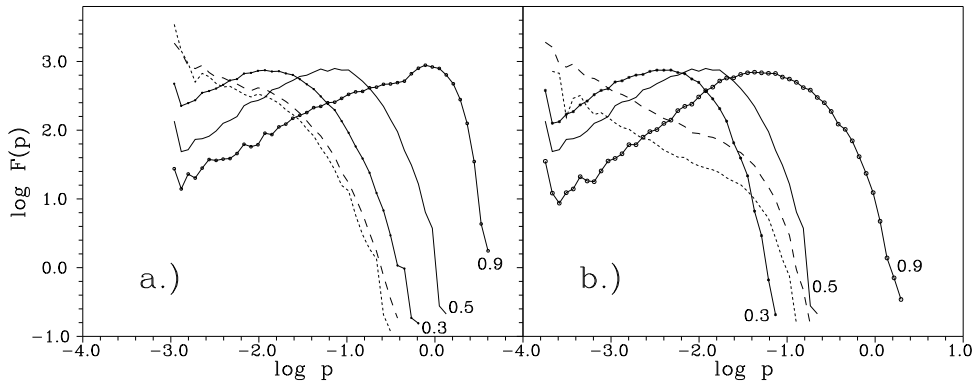}   
\vspace{0mm}
\caption{ 
Simulated particle distributions $F(p) \equiv dN(p)/d(\log  
p)$. Particles with the momentum $p = 1.0$ have a gyroradius equal to  
the jet radius, $R_j$ and the injection momentum is $p_0 << 1$. We take  
$R_{esc} = 2 R_j$ and $L_{esc} = R_j$. Full lines represent the spectra  
for particles injected at the distance of $1000 R_j$ upstream of the  
terminal shock; the respective jet velocities $U_1$ are given near these  
lines. The curve with long dashes denotes the spectrum for the terminal  
shock injection, while the one with shorter dashes presents the spectrum  
for the upstream injection with the side boundary acceleration neglected  
(these curves are for $U_1 = 0.5$). At the left panel (a) the results  
for the smaller cross-field diffusion $D = 0.0013$ are presented, while  
in the right panel (b) results for the nearly isotropic diffusion $D =  
0.97$  are given.} \end{figure}  
     
Spectra of particles accelerated at relativistic shock waves depend in a  
large extent on the poorly known physical conditions near the shock.  
Therefore, instead of attempting to reproduce a detailed shape of the  
particle spectrum let us, rather, consider a modification introduced to  
the {\it power-law with a cut-off} shock spectrum by the additional  
acceleration at the jet boundary (Ostrowski 1998). We neglect the  
radiation losses, i.e. the upper energy limit is fixed by the boundary  
conditions -- the finite shock and jet spatial extensions -- allowing  
for the escape of the highest energy particles. 
Let us consider the jet terminal shock resting  
with respect to the cocoon surrounding the jet. The  
conditions occurring behind the shock due to the flow divergence are  
modelled by imposing the particle free escape boundary at a finite  
distance, $L_{esc}$, downstream of the shock and in the cocoon adjoining  
this boundary. Also, another tube-like free escape boundary is  
introduced, surrounding the jet in a distance $R_{esc}$ from the jet  
axis. We assume the mean magnetic field to be parallel to the jet  
velocity both within the jet and in the cocoon. Possible significant  
deviations from this simple structure are partly accommodated in the  
simulations by also considering cases with unrealistically large  
cross-field diffusion. In the presented examples we take a ratio $D$ of  
the cross-field diffusion coefficient to the parallel diffusion  
coefficient to be $0.0013$ or $0.97$.  
     
In Fig.~8 one can consider spectra of particles escaping through the  
boundaries for the seed particle injection far upstream of the terminal  
shock and -- for $U_1 = 0.5$ -- compare these spectra to the ones  
generated at the shock. One may note that the distribution of particles  
accelerated at the jet side-boundary is very flat. This feature results  
from the character of the acceleration process with particles having  
a chance to meet the accelerating surface again and again due to  
inefficient diffusive escape to the sides. Contrary to that, the shock  
acceleration process determines the spectrum inclination due to the  
joint action of the particle energization at the shock and the  
continuous escape due to particle advection with the plasma. In the  
present simulations fixed spatial distances to the escape boundaries are  
assumed. Thus the escape probability grows with the particle momentum  
providing an energy cut-off in the spectrum. For the shock spectrum  
(Fig.~8b), in the range of particle energies directly preceding the  
cut-off, the spectrum exhibits some flattening with respect to the  
inclination expected for the standard shock acceleration mechanism.  
There are two reasons for that flattening: additional particle transport  
from the downstream shock region to the upstream one through the cocoon  
surrounding the jet and inclusion of the very flat spectral component  
resulting from the side boundary acceleration.  
     
\subsection{Large Lorentz factor jets}  
     
The expected in close vicinity of AGNs highly relativistic jets provide  
an exceptionally promising sites for accelerating particles. The only  
requirement is that some sufficiently energetic, with $\lambda_n \sim  
d$, seed particles exist in the jet vicinity. As even single  
interactions of such particles with the jet may boost its' energies on  
large factors (cf. Eq.~4.3), a rapid build-up of the cosmic ray energy  
density may result, with the bulk of energy contained in highest energy  
particles. If these particles become dynamically important one could  
speculate about the intermittent behaviour of the generated cosmic ray  
population providing the time modulation of the jet structure and the  
produced high energy photon field.  
  
\subsection{Shear layers near the black hole accretion disks}  
  
The same mechanism can work in all conditions allowing for large  
velocity gradients. One such suggested possibility arises in the vicinity of  
the accretion disk near the accreting black hole. A recent work of  
Subramanian et al. (1998) proposed this process to provide the energy  
source for ejection of the high Lorentz factor jets from vicinity of the  
quasar central engine.  
  
\section{Final remarks}  
  
The work done to date on the {\it test particle} cosmic ray acceleration  
at mildly relativistic shocks yielded not too promising results for  
meaningful modelling of the observed astrophysical sources. The main  
reason for that deficiency is -- in contrast to non-relativistic shocks  
-- a direct dependence of the derived spectra on the conditions at the  
shock. Not only the shock compression ratio, but also other parameters,  
like the mean inclination of the magnetic field or the wave spectrum  
shape and amplitude, are significant here. Depending on the actual  
conditions one may obtain spectral indices as flat as $\alpha = 3.0$  
($\gamma = 0.0$) or very steep ones with $\alpha > 5.0$ ($\gamma >  
1.0$). The background conditions leading to the very flat spectra are  
probably subject to some instabilities; however, there is no detailed  
derivation describing the instability growth and the resulting cosmic  
ray spectrum modification. The situation may become simpler for  
large $\Gamma$ shocks, where -- in the simulations of Bednarz \&  
Ostrowski (1998) -- the spectral index converges to the universal limit  
$\sigma_\infty \approx 2.2$.  
  
A true progress in modelling particle acceleration in actual sources  
requires a full plasma non-linear description, including feedback of  
accelerated particles at the turbulent wave fields near the shock wave,  
flow modification caused by the cosmic rays' plasma pre-shock  
compression and, of course, the appropriate boundary conditions. A  
simple approach to the parallel shock case was presented by Baring \&  
Kirk (1990), who found that relativistic shocks could be very efficient  
accelerators. However, it seems to us that in a more general case it  
will be very difficult to make any substantial progress in that matter.  
For very flat particle spectra the non-linear acceleration picture  
depends to a large extent on the detailed knowledge of the background  
and boundary conditions in the scales relevant for particles near the  
upper energy cut-off. The existence of stationary solutions is doubtful  
in this case. An important step toward considering detailed physics of  
the acceleration provides the work of Arons and  
collaborators described in Section 3.5, applicable in ultra-relativistic shocks.  
  
One may note that observations of possible sites of relativistic shock  
waves (knots and hot spots in extragalactic radio sources), which allow  
for the determination of the energetic electron spectra, often yield  
particle spectral indices close to $\alpha = 4.0$ ($\gamma = 0.5$). In  
order to overcome difficulties in accounting for these data Ostrowski  
(1994) proposed an additional {\it `law of nature'} for non-linear  
cosmic ray accelerators. The particles within different energy ranges do  
not couple directly with each other and are supposed to form independent  
`degrees of freedom' in the system. Our `law' provides that nature  
prefers energy equipartition between such degrees of freedom, yielding  
the spectra with $\alpha \approx 4.0$~.  
  
The acceleration processes at astrophysical shear layers may provide a  
viable explanation for some `strange' observational data. For example it  
may have significant consequences for the relativistic jet structure,  
particularly on the sub-parsec scales, and on the high energy photon  
radiation fields of AGNs (Ostrowski, in preparation). Also, these  
mechanisms could be responsible for accelerating particles up to the  
ultra high energies above EeV scales (Ostrowski 1998). An application of  
the viscous acceleration for the shear layer near the accretion disc in  
the AGN centre was presented by Subramanian et al. (1998). The discussed  
processes are particularly interesting because a simple inspection does  
not reveal any physical obstacles which could make them inefficient.  
Unfortunately, the resulting electron spectra depend to a large extent  
on the poorly known background physical conditions.  
     
\acknowledgements  
The present work was supported by the {\it Komitet Bada\'n Naukowych}  
through the grant PB 179/P03/96/11.

\end{document}